%% file: main.tex
\documentclass[conference]{IEEEtran}

\usepackage{graphicx}
\graphicspath{{img/}}

\usepackage{amsmath}

\usepackage[autostyle=false,style=american]{csquotes}
\MakeOuterQuote{"}

\usepackage[ruled]{algorithm2e}

\usepackage[caption=false,font=scriptsize,labelfont=sf,textfont=sf]{subfig}
\newcommand{\HH}{Hodgkin-Huxley}

\begin{document}
%
\title{Towards an Hybrid Hodgkin-Huxley \\ Action Potential Generation Model}

\author{\IEEEauthorblockN{Lautaro Estienne}
\IEEEauthorblockA{Universidad de Buenos Aires\\
Facultad de Ingeniería\\
Buenos Aires, Argentina\\
Email: lestienne@fi.uba.ar}
}


%


\maketitle

\begin{abstract}

Mathematical models for the generation of the action potential can improve the understanding of physiological mechanisms that are consequence of the electrical activity in neurons. In such models, some equations involving empirically obtained functions of the membrane potential are usually defined. The best known of these models, the \HH{} model, is an example of this paradigm since it defines the conductances of ion channels in terms of the opening and closing rates of each type of gate present in the channels. These functions need to be derived from laboratory measurements that are often very expensive and produce little data because they involve a time-space-independent measurement of the voltage in a single channel of the cell membrane. In this work, we investigate the possibility of finding the \HH{} model's parametric functions using only two simple measurements (the membrane voltage as a function of time and the injected current that triggered that voltage) and applying Deep Learning methods to estimate these functions. This would result in an hybrid model of the action potential generation composed by the original \HH{} equations and an Artificial Neural Network that requires a small set of easy-to-perform measurements to be trained.
Experiments were carried out using data generated from the original \HH{} model, and results show that a simple two-layer artificial neural network (ANN) architecture trained on a minimal amount of data can learn to model some of the fundamental proprieties of the action potential generation by estimating the model's rate functions. 

\end{abstract}

%
\IEEEpeerreviewmaketitle

\vspace{-.5em}
\section{Introduction}\label{sec:intro}

\input{intro.tex}
\section{Methods and Materials}\label{sec:met_mat}
\input{met_mat.tex}

\section{Results-Discussion}\label{sec:results_discussion}
\input{results.tex}

\section{Conclusions}\label{sec:conclusion}
\input{conclusion.tex}
\end{document}

%% file: intro.tex
\begin{figure*}[!t]
	\centering
	\subfloat[]{\includegraphics[width=1.5in,height=3cm]{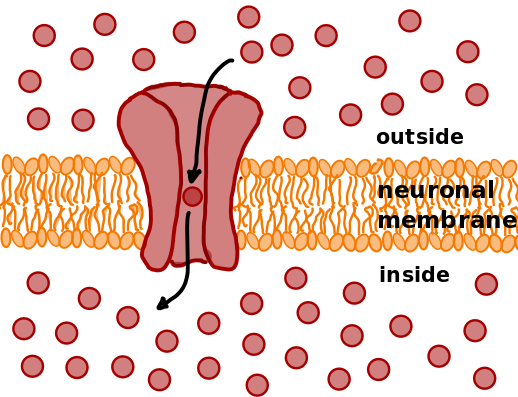}
	\label{fig:ion_channel}}
	\hfil	
	\subfloat[]{\includegraphics[width=2.5in,height=3cm]{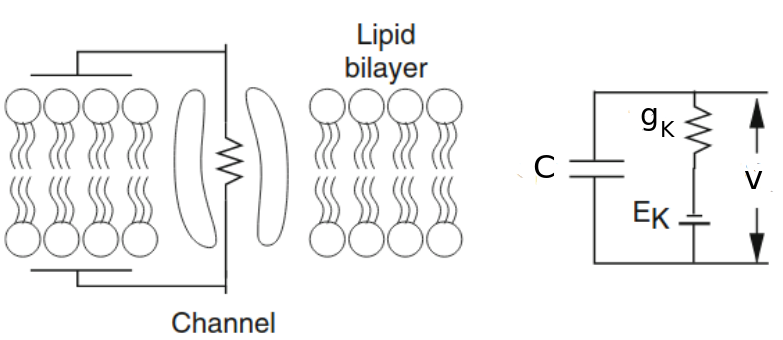}
	\label{fig:ion_circuit}}
	\hfil
	\subfloat[]{\includegraphics[width=2.5in,height=3cm]{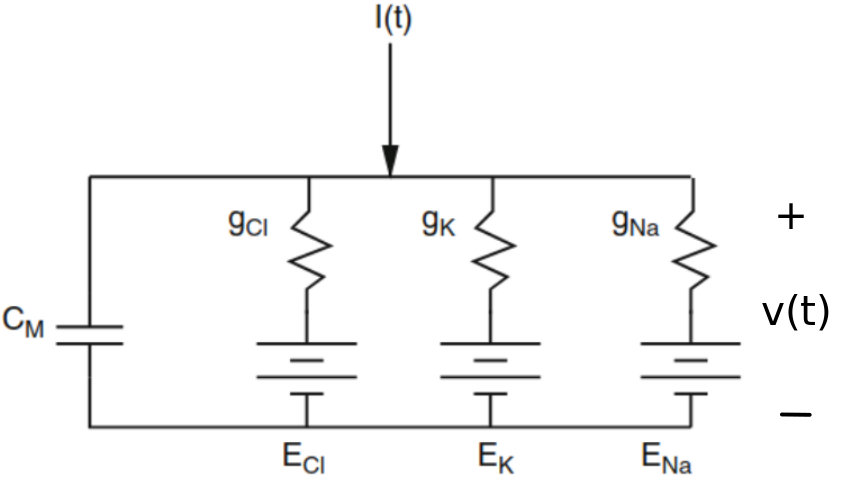}
	\label{fig:circuit_full}}
	\caption{\ref{fig:ion_channel} Representation of an open ion channel. These channels allow selected ions to move, via diffusion, from one side of the membrane to the other. \ref{fig:ion_circuit} Electrical model of a cell membrane made up of potassium channel only. \ref{fig:circuit_full} \HH{}'s electrical model of the neuron membrane.}
	\label{fig_sim}
	\vspace{-1em}
\end{figure*}

The cell membrane provides a structure capable of protecting the cell from its external environment and maintaining at the same time selective communication with it. As a result, the cell can maintain internal work without interference from the external environment and generate signals between cells through ion flow. The cell membrane accomplishes all this work thanks to a structure consisting in a phospholipid \emph{bilayer}, where aggregates of globular proteins and protein-lined pores (called \emph{channels}) are irregularly distributed. This structure allows molecules present in both intracellular and extracellular environments to be transported across the membrane. In that way, sodium and potassium ions passing through ion-specific channels as a consequence of electrical forces interactions is an example of a process called \emph{simple diffusion}, which does not involve energy expenditure from the cell (figure \ref{fig:ion_channel}). In addition, concentration differences are set up and maintained by active mechanisms (i. e. that use energy) to pump ions against their concentration gradient \cite{purves}. 
As a result of all these transport mechanisms, a controlled variation of charge concentrations between the inside and outside of the cell occurs in some specialized cells such as \emph{neurons} (nervous tissue) or \emph{myocytes} (muscle tissue). This leads to a resulting change in the potential difference on both sides of the membrane that can be used as a complex signaling process with other cells. In the case of neurons, the membrane contains specific ionic channels, mostly $K^+$ (potassium), $Na^+$ (sodium), and $Cl^-$ (chlorine), that allow simple diffusion transport of only one type of these ions through them. Furthermore, the opening and closing of these channels strongly depend on the voltage across the membrane, producing a control mechanism for the input and output of ions. In addition, $Na^+ \!-\! K^+$ pumps that regulate the excess of ions inside the cell are also part of this control mechanism.

As a result of this structure present in all neurons, the membrane generates an impulsive change in the voltage across the membrane when an initial stimulus occurs. This change is usually known as \emph{action potential}. The electrical activity of neurons is measured in terms of the number of action potentials that it generates per unit of time and is responsible for the most complex physiological processes known, such as the senses, motor skills, reasoning, and language. For this reason, it is of particular interest to study the process of generation of the action potential in the membrane and its modeling.

One of the best-known action potential generation models is the model developed by Hodgkin and Huxley \cite{HH1952}, who not only made a quantitative and precise description of the action potential generation but also laid the foundations to understand a large part of the behavior of neurons. In their model, covered with more detail in section 2, each ion channel contains a set of four gates which follows a probabilistic opening and closing dynamics. This dynamic is governed by two mathematical functions (the opening and closing \emph{rates}) that depend on the membrane voltage. These rates have to be empirically determined, and some popular techniques such as the \emph{patch-clamp} and its variations \cite{clampdata} are normally used to achieve this goal. That way, a parametric function is usually proposed to match the rate's properties and its parameters are determined through measurements of the current and voltage in the channel. However, the way these measurements are usually carried out involves a number of constraints that must be held to correctly determine the function's parameters \cite{patchcond,mathphy,peskin}. For example, recordings of the current passing through the channel must be done for a few or even a single channel because rate functions are different for different types of channels. In addition, the current must be recorded when a small known voltage step is applied to the membrane in the channel proximity. If a large voltage step is applied, the effect of the membrane capacitance comes into play and a portion of the recorded current does not correspond to ions passing through the channel, but to those accumulating on the membrane surface. Also, voltage differences that might develop between one location and another along the membrane should be minimized to ensure that the membrane potential is a function only of time, and does not varies with the position along the membrane too. This last condition was achieved by Hodgkin and Huxley in their squid giant axon experiments and it was one of the major reasons for their success \cite{hh2}.

Motivated by these limitations, we propose to adapt the \HH{} original model so that the rate functions mentioned above can be obtained using two simple measurements: the membrane voltage in a small part of its surface (but not restricted to a single channel) as a function of time and the injected current that triggered that voltage. 
Also motivated by the previous success of Deep Learning in some biomedical signal processing fields like Speech or ECG Processing \cite{ecgdl,speechdl}, in this paper we show that it is possible to use only these measurements to train the model rate functions with Artificial Neural Networks (ANN) and Gradient Descent-based algorithms \cite{adam}. We propose this method as a starting point to advance towards an hybrid action potential generation model. This kind of model takes advantage of the deep learning methods and keeps at the same time a physiologically plausible interpretation of its internal working.

Previous studies have been carried out concerning the parameter estimation of \HH{} from biological recordings, and some methods including gradient-descent \cite{gradientdescent}, metaheuristics \cite{metaheuristics} and self-organizing state-space modeling \cite{statespace} have been proposed. In these methods, the ultimate goal is to find the best estimation of the model parameters. Another approach consisting in defining a deep learning based meta-model of the original \HH{} model has been proposed by \cite{metamodel}. Here, a Deep Neural Network is trained from the original model to improve its predictions, but additional interpretation of the network's insight has to be done in order to extract biological information. As far as we know, the approach proposed in this work has never been explored.

%% file: met_mat.tex
\subsection{\HH{} model}

In 1952 Hodgkin and Huxley \cite{HH1952} proposed an action potential generation model from the study of the squid giant axon. In this model, an equivalence is made between the time response of the voltage measured across de cell membrane and the time response of an electrical circuit consisting of a capacitor in parallel with three conductances (figures \ref{fig:ion_circuit} and \ref{fig:circuit_full}). The first two correspond to the conductance of the $Na^+$ and $K^+$ channels, respectively, which depend on the voltage measured across the membrane. The third conductance, called \emph{leakage}, is proportional to the flow of ions that do not leave the cell and does not depend on the applied voltage. In addition, progressive accumulation of ions on both sides of the membrane produces an electrical current through the membrane, hence a capacitance $C$ is included in the model. 
Finally, the model includes three resting potentials $E_{Na}$, $E_{K}$ and $E_{L}$, which are a consequence of the fact that, in the absence of external stimuli, there is a difference in the intracellular and extracellular ion concentrations. Although these concentrations change during charge flowing, changes provoked in these potentials by the current flow can be considered negligible, so rest potentials are treated as constants.

As a result, when a current $i(t)$ in applied to the membrane through an electrode or because a synaptic process occurs near the channels the following equation must hold:
\begin{equation}
	i(t) \!=\! {\scriptstyle C} \frac{dv\!{\scriptstyle (t)}}{dt} \!+\! g_{\scriptscriptstyle \! N \! a}\!{\scriptstyle (t)} \! (v\!{\scriptstyle (t)}\!-\!{\scriptstyle E}_{\scriptscriptstyle \! N \! a}) \!+\! g_{\scriptscriptstyle \! K}\!{\scriptstyle (t)} \! (v\!{\scriptstyle (t)}\!-\!{\scriptstyle E}_{\scriptscriptstyle \! K}) \!+\! g_{\scriptscriptstyle \! L}(v\!{\scriptstyle (t)}\!-\!{\scriptstyle E}_{\scriptscriptstyle \! L}) 
\end{equation}
which can be rewritten as
\begin{equation}\label{eq:memb_dif_orig}
	i(t) = C\frac{dv(t)}{dt} + g(t) (v(t) - E(t))
\end{equation}
where
\vspace{-1em}
\begin{align*}
	g(t) &= g_{Na}(t) + g_{K}(t) + g_L \\
	E(t) &= \frac{E_{Na} g_{Na}(t) + E_{K} g_{K}(t) + E_{L} g_L}{g_{Na}(t) + g_{K}(t) + g_L}
\end{align*}

Although this membrane model allows us to quantitatively and qualitatively visualize its behavior, the great contribution of the \HH{} model was the measurement of the conductance of each channel separately. As described in section \ref{sec:intro}, this could be done because the conditions of the experiments established by Hodgkin and Huxley in the squid giant axon allowed them to obtain the current flowing through a particular type of channel when a known constant voltage is applied to it. With these measurements, a representation of the conductances $g_{Na}$, $g_{K}$ and $g_{L}$ could be done, completing the model described so far. This representation is described next.

For both $Na^+$ and $K^+$, the channels are made up of 4 internal gates. The channel will only be open if all four gates are open. Moreover, the four gates could operate differently from one channel to the other, but always independently of each other. In the case of $K^+$, the four gates are identical, in the sense that the opening probability (which we will call $n$) is the same for each gate. Conversely, in the case of $Na^+$, the channel has four gates but three of them have an opening probability $m$, while the remaining one has a probability $h$, and they behave differently from each other. 
In addition, the dynamic of the opening probability $s(t), s \in \{n,m,h\}$ responds to the following diffential equation:
\vspace{-.2em}
\begin{align}
	\frac{ds(t)}{dt} &= \alpha_s(v(t)) (1-s(t)) - \beta_s(v(t)) s(t) \label{eq:s_diff_orig}
\end{align}
\vspace{-.2em}
In each case, $\alpha_s(v)$ and $\beta_s(v)$ are, respectively, the opening and closing rates for the $s(t)$ probability. These functions depend solely on the voltage applied to the membrane and are those obtained by Hodgkin and Huxley from their experimental measurements. 
Since gates are independent of each other, the total conductance of each channel is proportional to the average number of open channels at a given time. This leads to the following equations:
\vspace{-.2em}
\begin{align}
	g_{Na}(t) &= \bar{g}_{Na} m^3(t) h(t) \label{eq:g_na}\\
	g_{K}(t) &= \bar{g}_{K} n^4(t) \label{eq:g_k} 
\end{align}
\vspace{-.2em}
where $\bar{g}_{Na}$ and $\bar{g}_{K}$ are empirically obtained values. With this definitions, the model is complete and can be tested against the one presented in this paper.

\subsection{Training the \HH{} Model.}\label{sec:mat_met_train}

In this work we propose that the rates functions $\alpha_n(v)$, $\beta_n(v)$, $\alpha_m(v)$, $\beta_m(v)$, $\alpha_h(v)$ and $\beta_h(v)$ can be redefined and trained using samples of the $i(t)$ current and the $v(t)$ voltage present in equation (\ref{eq:memb_dif_orig}). In this way, there is no need to perform other measurements that the voltage across the membrane and the current that flows in its proximity. We assumed for simplicity that the constants $\bar{g}_{Na}$, $\bar{g}_{K}$, $g_{L}$, $E_{Na}$, $E_{K}$, $E_{L}$ and $C$ are always provided because we also want to study whether the model can learn a rate function that follows the shape of an Artificial Neural Network. However, further works could consider these constants as part of the parameters to be trained and the algorithm would be the same.

In order to train the model, discrete time versions of equations (\ref{eq:memb_dif_orig}) and (\ref{eq:s_diff_orig}), are obtained by first order approximation:
\begin{align}
	i(t) &= C\frac{v(t) \!-\! v(t \!-\! \Delta t)}{\Delta t} + g(t) (v(t) \!-\! E(t)) \label{eq:memb_diff_num} 
	\\
	\frac{s(t) \!\!-\!\! s(t\!\!-\!\!\Delta \! t)}{\Delta \! t} &= 
		\alpha_s(v(t \!\!-\!\! \Delta \! t)) (1\!\!-\!\!s(t)) \!- \! \beta_s\!(v(t\!\!-\!\!\Delta \! t)) s(t) \label{eq:s_diff_num} 
\end{align}
where $s \in \{n,m,h\}$. Rearranging these equations, an iterative algorithm to predict the voltage $v(t)$ as a function of $i(t)$, $v(t-\Delta t)$ and the probabilities $n(t)$, $m(t)$ and $h(t)$ is obtained:
\begin{align}
	v(t) &= \frac{v(t-\Delta t) + \frac{\Delta t}{C} (g(t) E(t) + i(t))}{1 + \frac{\Delta t}{C} g(t)} \label{eq:v_update} \\
	s(t) & = \frac{s(t-\Delta t) + \Delta t\; \alpha_s(v(t-\Delta t))}{1 + \Delta t \left( \alpha_s(v(t-\Delta t)) + \beta_s(v(t-\Delta t)) \right)} \label{eq:s_update}
\end{align}
These equations provide a way of calculating the value of $v(t)$ from $v(t-\Delta t)$ because $g(t)$ and $E(t)$ depend on probabilities $n(t)$, $m(t)$ and $h(t)$, which at the same time depend on $v(t-\Delta t)$ and their values at the previous step. If the membrane was at rest at some value $v_{hold}$ before some time $t_0$, i. e. $v(t)=v_{hold}$ for $t < t_0$, it can be assumed that $s(t) \approx s(t - \Delta t)$ for $t < t_0$, and the initial values for $s(t)$ can be calculated:
\begin{equation}
	s(t_0) \approx \frac{\alpha_s(v(t-\Delta t))}{\alpha_s(v(t-\Delta t)) + \beta_s(v(t-\Delta t))}\label{eq:s_init}
\end{equation}
If the rate functions $\alpha_s(v)$ and $\beta_s(v)$ for $s \in \{ n,m,h\}$ are provided, the model is completely defined by equations (\ref{eq:memb_diff_num}), (\ref{eq:s_update}) and (\ref{eq:s_init}). Next, we discuss how this rates can be defined in order to find a model that can be trained from a set of sequence pairs $X_{train} = \{(i_{1_n},\allowbreak v_{1_n}), \ldots, (i_{T_n},v_{T_n})\}_{n=1}^N$. Each of this sequence pairs are time samples of the current $i(t)$ and the voltage $v(t)$ using a sample period $\Delta t$. 

For the experiments of this paper we took into account the monotony of these functions, and designed 6 Artificial Neural Networks that follows a 2-layer architecture with a single neuron in each layer:

\vspace{-1em}
{\small
\begin{align*}
	\alpha_n(v) &= f_{\alpha_n}(v,{\boldsymbol\theta_{\alpha_n}}) = f_1\left( -f_2(-w_{1\alpha_n} v + b_{\alpha_n}) w_{2\alpha_n} + b_{2\alpha_n} \right) \\
	\beta_n(v) &= f_{\beta_n}(v,{\boldsymbol\theta_{\beta_n}}) = f_1\left( -f_2(w_{1\beta_n} v + b_{\beta_n}) w_{2\beta_n} + b_{2\beta_n} \right) \\
	\alpha_m(v) &= f_{\alpha_m}(v,{\boldsymbol\theta_{\alpha_m}}) = f_1\left( -f_2(-w_{1\alpha_m} v + b_{\alpha_m}) w_{2\alpha_m} + b_{2\alpha_m} \right) \\
	\beta_m(v) &= f_{\beta_m}(v,{\boldsymbol\theta_{\beta_m}}) = f_1\left( -f_2(w_{1\beta_m} v + b_{\beta_m}) w_{2\beta_m} + b_{2\beta_m} \right) \\
	\alpha_h(v) &= f_{\alpha_h}(v,{\boldsymbol\theta_{\alpha_h}}) = f_1\left( f_3(-w_{1\alpha_h} v + b_{\alpha_h}) w_{2\alpha_h} + b_{2\alpha_h} \right) \\
	\beta_h(v) &= f_{\beta_h}(v,{\boldsymbol\theta_{\beta_h}}) = f_1\left( f_3(w_{1\beta_h} v + b_{\beta_h}) w_{2\beta_h} + b_{2\beta_h} \right)
\end{align*}
}%
where $f_1(x)=\max\{0,x\}$, $f_2(x)=\log\sigma(x)$, $f_3(x)=\sigma(x)$ and $\sigma(x)$ is the sigmoid activation function. To each of these functions correspond a set of parameters ${\boldsymbol\theta_{ij}}, i \in \{ \alpha, \beta\}, j \in \{ n,m,h\}$ to be estimated by the algorithm. In the above equations, we have replace ${\boldsymbol \theta_{ij}}$ by $\{w_{1ij},b_{1ij},w_{2ij},b_{2ij}\}$ to represent each set of parameters separately. These architectures where defined in order to keep the fact that $\alpha_n(v)$, $\alpha_m(v)$ and $\beta_h(v)$ are increasing functions of the voltage while $\beta_n(v)$, $\beta_m(v)$ and $\alpha_h(v)$ are decreasing, and that all rates must be positive. 

The fact that the rate functions can be replaced with an ANN and trained using a simple measurement could represent the starting point of an hybrid action potential generation model that can take the advantages of a Deep Learning model's ability to learn from large amount of data, while keeping a physiologically plausible interpretation of the action potential generation.

\subsection{Methods}

The following sections will show the experiments carried out for the original model compared to the one trained with the proposed method. The goal of these sections will be to show that the trained model keeps some of the fundamental properties of the nerve action potential. 

To simulate the original model, equations (\ref{eq:v_update}), (\ref{eq:s_update}) and (\ref{eq:s_init}) were implemented using \texttt{Python}'s numerical library \texttt{Numpy}. Values for $v_{hold}$ and $\Delta t$ were fixed to $-70\mathrm{mV}$ and $0.05\mathrm{ms}$ for all simulations, and the rates and constants used for this part were the one used in \cite{peskin}.

To train and validate the proposed model, a train set $X_{train}$, a validation set $X_{val}$ and a test set $X_{test}$ of sequence pairs were built from the original model. Each sequence pair consisted of a single-impulse-shaped current input signal and its corresponding original model's voltage output. The height of the current impulses was kept in the range of $0\;\mathrm{mA}$ to $50\;\mathrm{mA}$ to cover the two states of the neuron (firing and resting) and random noise ($SNR=80\;\mathrm{dB}$ at the output) was added to all signals for simulation purposes. This dataset contained only 2 pairs of signals for training, 10 for validation and 10 for testing. Table \ref{tab:splits} shows the heights $ip$ of the impulses used in each split, and it can be notice from figure \ref{fig:threshold} that all input signals in the training set generate an action potential, whereas some impulses in the validation and test sets do not. 
Our goal was to keep de number of training signals as low as possible to reduce the number of measurements required for the model to be successfully trained. To increase the number of training samples some data augmentation techniques were implemented on the train split of the dataset, including random noise addition and random time-shifting. 

Some of the most used gradient descent-based algorithms were tested for the training step, but the Adam optimizer \cite{adam} was the one with the best accuracy and training time. The cost function used to train the model was the $L_1$ loss, that corresponds to the absolute value of the point-to-point difference between the predicted and the true signal. This cost function was chosen to obtain a better visual similarity between the two signals and to minimize the effect that this non linear system may cause in, for example, two signals that has low mean square error but not absolute error. Simulations from this part were coded using \texttt{Pytorch}, and training was performed using the \texttt{autograd} module).

\begin{table}
	\centering
	\caption{}\label{tab:splits}
	\vspace{-1em}
	\begin{tabular*}{.5\textwidth}[!t]{l|c}
		Dataset split & Height of the current impulse [$\mu\mathrm{A}$]\\ \hline
		\raisebox{-.5\height}{$X_{train}$ ($N=2$)} & \raisebox{-.5\height}{10.0, 20.0} \\[.5em]\hline
		\raisebox{-.5\height}{$X_{val}$ ($N=10$)} & \raisebox{-.5\height}{0.0, 0.5, 1.0, 2.0, 4.0, 8.0, 11.0, 21.0, 35.0, 50.0} \\[.5em]\hline
		\raisebox{-.5\height}{$X_{test}$ ($N=10$)} & \raisebox{-.5\height}{1.2, 2.1, 3.4, 4.6, 7.6, 9.1, 13.1, 17.9, 27.4, 30.5}
	\end{tabular*}
	\vspace{-1.5em}
\end{table}

Regarding the physiological properties of the neuron membrane, one of the most important characteristics of a neuron's action potential is its "all-or-none" behavior, in the sense that there is a \emph{threshold} voltage above which the neuron will be activated. If the voltage reaches this value, the membrane completes its cycle of totally depolarizing, then hyperpolarizing and finally reaching rest. Otherwise, the membrane depolarizes a certain level but it returns to rest state without going through the entire process. 
Another important property of the action potential is its \emph{refractory time}. If an action potential has been generated, sodium channels inactivate, and a certain time (called \emph{absolute refractory time}) must pass before they can be opened again, even if another impulse of the same characteristics is injected in this time interval. Moreover, even if a second impulse is injected after the absolute refractory time, there is a period (called \emph{relative refractory time}) where an action potential can be generated but only with a stronger stimulus than the first one.

These two properties of the action potential were tested using the same impulse currents as inputs for the two models (the original one and the one trained with gradient descent) and monitoring the membrane potential voltage generated for each of them. Additionally, we tested the response of the model to a non-localized current input, which may be the case of a stimulus generated in some neurons such as the sensory ones. To do that, we record the period of the resulting action potential as a function of the constant current value injected at the model's input. 

Finally, the time response of the potassium and sodium conductances to a step-change in the membrane voltage was also simulated for both models, given that it can be used to provide more evidence of the learning capacity of the model. When a step-change in the membrane voltage from a value near rest to a higher value is applied, potassium conductance slowly increases until it reaches a maximum value, whereas the sodium conductance increases faster but it rapidly turns back to its initial value because these channels inactivate. This behavior is included in the \HH{} model via the $n$, $m$ and $h$ probabilities and their dynamics. This can confirm that the rates functions to be trained are involved in all of these model's properties.

%% file: results.tex
\begin{figure*}[!t]
	\centering
	\begin{minipage}[t]{.30\textwidth}
		\subfloat[]{\includegraphics[width=\textwidth]{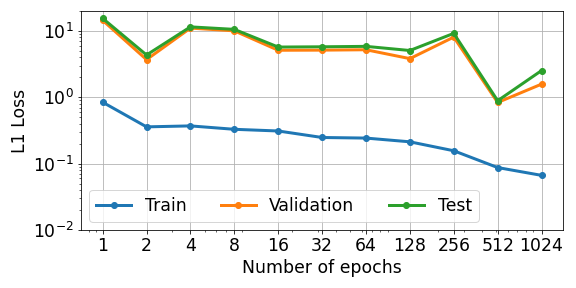}\label{fig:train}}\\
		\subfloat[]{\includegraphics[width=\textwidth]{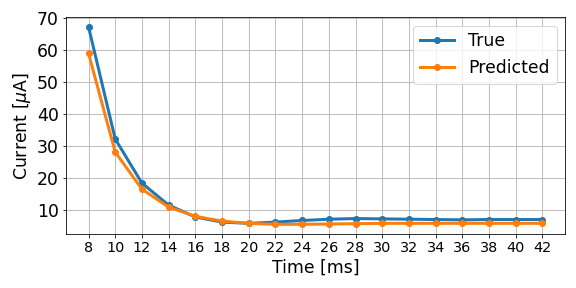}\label{fig:curr_vs_time}}\\
		\subfloat[]{\includegraphics[width=\textwidth]{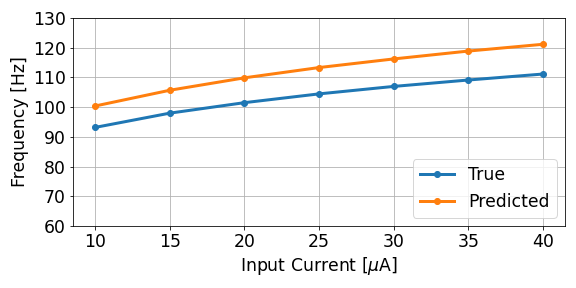}\label{fig:freq}}
	\end{minipage}
	\hfil
	\begin{minipage}[t]{.50\textwidth}
		\subfloat[]{
			\includegraphics[width=\textwidth,height=3.8cm]{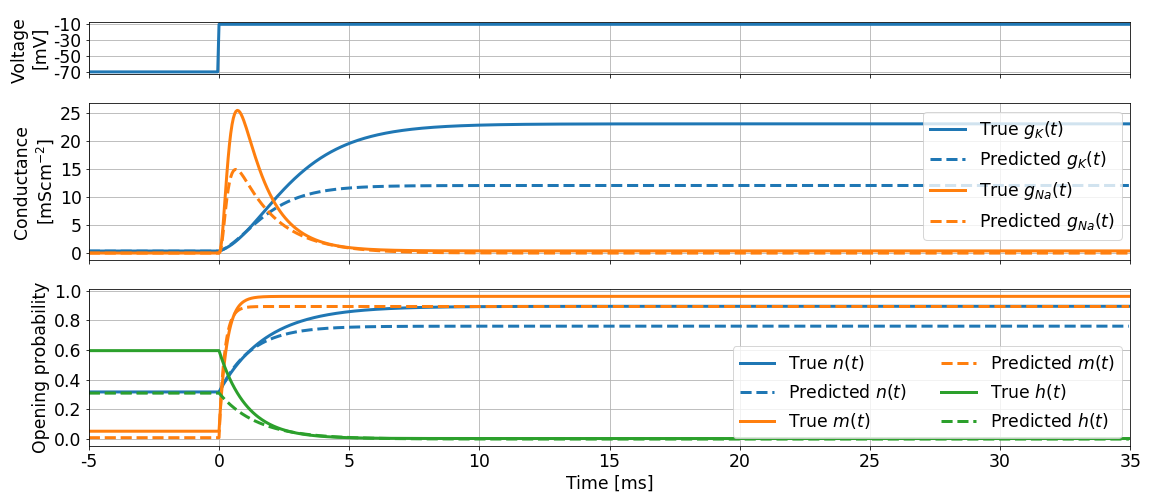}
		\label{fig:conductance}}\\[0pt]
		\subfloat[]{
			\includegraphics[width=\textwidth,height=5cm]{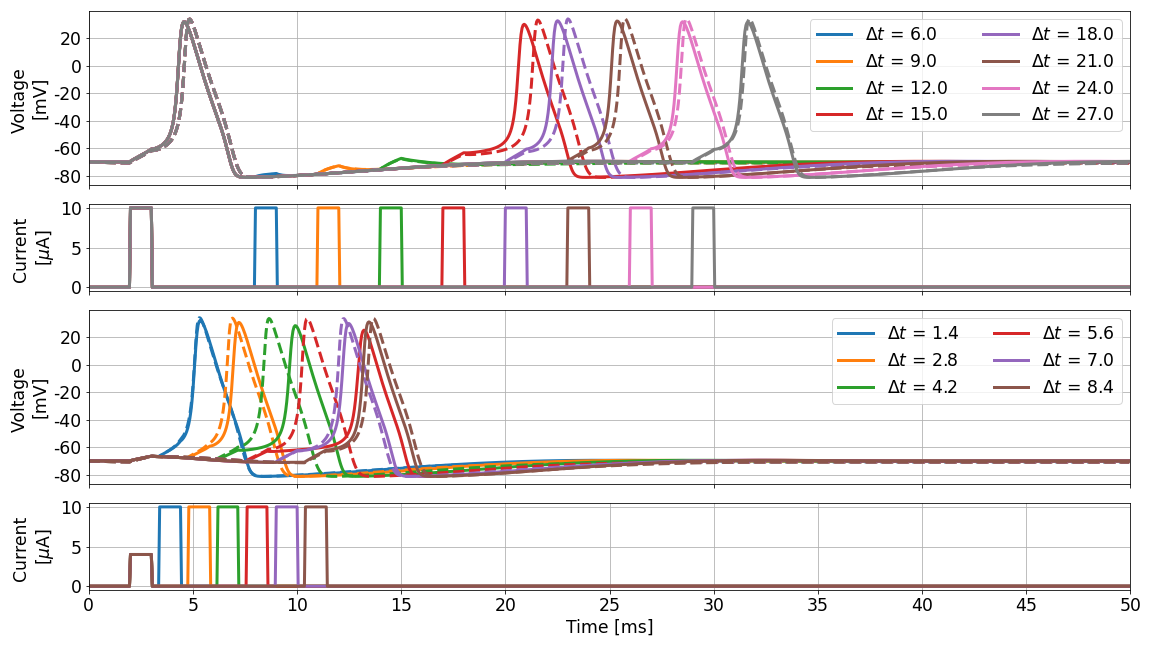}
		\label{fig:time}}\\[-5pt]
	\end{minipage}
	\subfloat[]{\includegraphics[width=.90\textwidth,height=3.3cm]{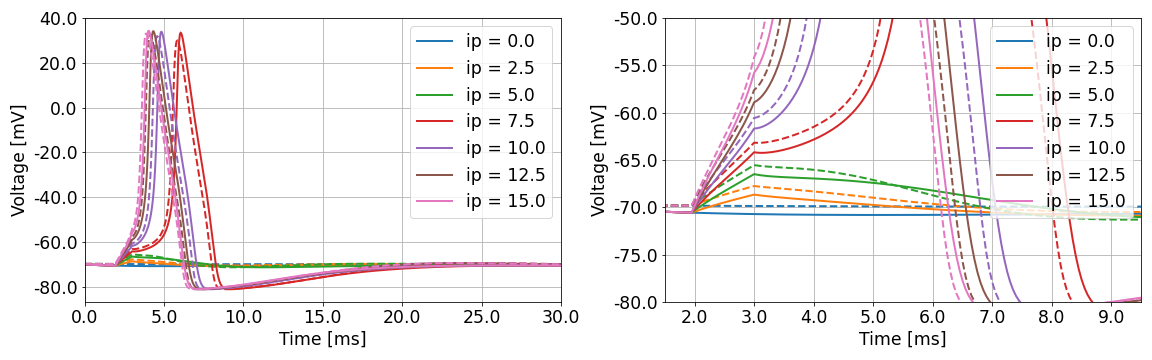}\label{fig:threshold}}\\[-5pt]
	\caption{\ref{fig:train} $L_1$ loss as a function of the number of epochs for train, validation and test splits. \ref{fig:curr_vs_time} Minimum current height of the second impulse needed to generate an action potential, as a function of the time difference between the two stimulus. \ref{fig:freq} Frequency of the output as a function of the current input. \ref{fig:conductance} Conductances and opening probabilities response to a voltage step change. \ref{fig:time} Time response to a two stimuli input. \ref{fig:threshold} Voltage output for different current impulse input.}
	\label{fig:6}
	\vspace{-1em}
\end{figure*}


Figure \ref{fig:train} shows the score ($L_1$ loss) of the proposed model on the train, validation and test splits as a function of the number of epochs used to train it. In all cases, Adam optimization with no regularization was implemented using an augmented version of the train split and the model described in section \ref{sec:mat_met_train}. This new dataset consisted in 2048 samples obtained from the original set using random noise addition with a maximum standard deviation of $\sigma = 0.05$ and random time shifting. The size of the batch was fixed to 512 in all cases and the learning rate was chosen from a grid of values ranging from $1\mathrm{e}-3$ to $1\mathrm{e}-5$ and by calculating its score in the validation set. It was found that the best validation score was obtained for a learning rate of $0.005$ for a 1024-epochs training, and it was fixed to that value for the rest of the experiment. This was done to study the learning process of the model.

It can be noticed that with only two training signal pairs, $L_1$ error decreases with the number of epochs in all splits, indicating that the model is actually learning. What is even more surprising is that although the signals used for training are samples of the neuron's firing state, the model can learn to correctly reproduce a voltage response to a stimulus that is not enough to generate an action potential. This phenomenon can be appreciated in figure \ref{fig:threshold}, where the voltage response to a set of current impulses with different heights are plotted. The figure shows in solid line the original \HH{} response and in dashed line the response of a model trained with the same conditions described above, for a total of 1024 epochs. As anticipated, impulses with height $ip\leq 5.0$ that do not produce an action potential in the original model, do not generate an action potential in the trained model either, even that signals of this kind were never shown to the model during training.

However, some indicators of an early limitation in model's generalization capability could be found in figure \ref{fig:train}. For instance, the gap between these curves and the training one could indicate that the model is overfitting to the train set. In addition, validation and test loss curves present some oscillations, indicating that the cost function could present several local minima. Recalling the fact that the membrane voltage at rest is very different than a firing one, we hypothesize that these observations could also be a consequence of the difference between the train and test data generation distribution. In general, all trainings performed showed the same difficulty to reduce the validation error.


On the other hand, figures \ref{fig:curr_vs_time}, \ref{fig:freq} and \ref{fig:time} show the results of the experiments that were performed to study our model's capability to generate more than one action potential when exposed to a more complex stimulus than before. First, voltage response to a stimulus made of two current impulses separated by a time $\Delta t$, each of $1\mathrm{ms}$ duration, is presented (figures \ref{fig:time}-top and \ref{fig:time}-bottom). Figure \ref{fig:time}(top) shows the resulting action potentials when the heights of the impulses are both $ip=10\mathrm{\mu A}$. Although it was previously shown that a single current impulse with these properties could generate an action potential if applied to a resting membrane, this simulation shows that for $\Delta t \leq 12\mathrm{ms}$ the second impulse does not generate a second action potential in either of the models. Conversely, figure \ref{fig:time}(bottom) shows that if the first impulse is too low ($ip=4\mathrm{\mu A}$) to generate an action potential, the second action potential could be generated with the original impulse. 

While these experiments are consistent with the existence of an absolute and relative refractory period, we additionally perform a simulation that keeps track of the second action potential's threshold voltage when the time interval between the two pulses is increased. Because the voltage is harder to measure than the current in this simulation, we plotted in figure \ref{fig:curr_vs_time} the minimum $ip$ value needed for the second impulse to generate an action potential with both the original and the trained model. We found that for both models, the absolute refractory period lies in the interval $[6\mathrm{ms},8\mathrm{ms}]$ because no value of $ip$ that could generate a second action potential when $\Delta t = 6\mathrm{ms}$ was found for any of the models. Conversely, for $\Delta t > 8\mathrm{ms}$ a lower $ip$ is needed to generate a second action potential as the interval between the two stimuli increases, until it stabilizes for values larger than $36\mathrm{ms}$. This confirms that, once trained, the proposed model is capable of successfully reproduce the property of having a refractory period after an action potential has been generated. However, how well this is done requires a more accurate measurement and it is not covered in this work. 
To complement this analysis, we compared the response of both models to a constant current value. The resulting output signal followed in both cases the shape of a sequence of action potentials repeated periodically. The frequency of the signal was recorded for each model and plotted for comparison in the figure \ref{fig:freq}. While there is clearly a gap between the plotted curves, in both cases the frequency of the output increases with the current value at the input. A $20.1\%$ of average relative error was estimated using this measurements, which may be a consequence of the model's limitation to generalize to new data. However, the fact that the trained model demonstrate to generate a physiologically reasonable output from a completely new input is, without a doubt, an achievement of the work.

Finally, additional evidence that the temporal physiological behavior holds for the trained model is shown in figure \ref{fig:conductance}. Here we plotted the channel's response to a voltage step change from a value near rest to a more positive one. It is interesting to notice that the general properties of the channel conductances and the opening probability hold even that the model was trained only with current impulses as input and sharped voltage signals as output, and that no additional information about each channel was given as training data. As instance, the potassium conductance of the trained model follows the classic slow-increasing shape of the original model, while the sodium conductance holds its ability to rapidly increase after the voltage step but also to return to its resting value as time goes on. However, a numerical difference between the predicted and the original conductances can be appreciated in the simulation, probably as a result of the accumulated difference in the opening probabilities plotted in the plot below. We hypothesize that this difference could be minimized with more training signals and more regularization methods in the training phase, but such analysis is leaved for future works.

%% file: conclusion.tex
In this paper, we modify the \HH{} model with the aim of advance towards the implementation of an action potential generation model that can be trained from a large amount of easy-to-perform laboratory measurements while keeping at the same time a physiologically plausible interpretation of its equations. To achieve this, we propose to introduce some techniques used in the data science field of Deep Learning, such as Artificial Neural Networks and Gradient Descent optimization, obtaining an hybrid model. Results performed on data generated with the original \HH{} model show that the opening and closing rates can be re-defined to follow the shape of a two-layer Neural Network and trained on a minimal amount of data to model some of the most important physiological properties of the action potential. This could represent the starting point for defining a model that can adapt to a wide range of neuron's action potential, finding physiological patterns between them.